# Processing and Characterization of Multiferroic Bi-relaxors

**Ashok Kumar[1], R. S. Katiyar[1*], J. F. Scott[1,2]**

[1]Department of Physics and Institute for Functional Nanomaterials, University of Puerto Rico, San Juan, PR 00931-3343 USA

[2]Cavendish Laboratory, Dept. Physics, Cambridge University, Cambridge CB3 0HE, U. K.

**Abstract:** We compare chemical solution deposition (CSD), and pulsed-laser-deposition (PLD), specimens of the new room-temperature, single-phase, multiferroic magnetoelectric, $[PbFe_{2/3}W_{1/3}O_3]_x[PbZr_{0.53}Ti_{0.47}O_3]_{1-x}$ ($PZTFW_x$ ~ 0.40<x<0.20) with polarization, loss (<1%), and resistivity (typically $10^8$ ohm.cm) equal to or superior to $BiFeO_3$. Single phase polycrystalline multiferroics $PZTFW_x$ thin films were fabricated on platinized silicon substrate by CSD and as epitaxial single-crystal films on MgO substrate by PLD. High dielectric constants (1200- 3000), high polarization (30 - 60 $\mu C/cm^2$), weak saturation magnetization (0.48 - 4.53 $emu/cm^3$), a broad dielectric temperature peak, high-frequency dispersion, low dielectric loss and low leakage current were observed in these materials, suggesting the family as candidates for room-temperature multiferroic devices. The ferroelectric switching in these materials can be suppressed or quenched with applied magnetic field.

***Corresponding author**

---

Email: rkatiyar@uprrp.edu (Prof. Ram S. Katiyar) and Email: jfs32@hermes.cam.ac.uk (Prof. James F Scott)

## 1. Introduction:

In the last decade numerous works on oxide materials have been carried out to explore their multiferroicity. As emphasized by Hill [1], we know the simultaneous existence of ferroelectricty and ferromagnetism is statistically very unlikely in single phase materials: ferroelectricity results from relative shifts of negative and positive ions that induce surfaces charges and is favored by empty *d* orbitals in covalent oxides such as perovskite titanates; on the other hand, magnetism is related to ordering of spin of electrons in incomplete electron shells, resulting from partially filled *d* orbitals. Most of the existing room temperature single-phase multiferroics don't posses linear magnetoelectric (ME) coupling (a bilinear term PM in the free energy) due to symmetry restrictions. Smolenskii. et al. had discovered various multiferroics compounds in the late 1950s, among which $Pb(Fe_{0.66}W_{0.33})O_3$ (PFW) was one of the promising candidates, having a ferroelectric relaxor transition near 180 K and an antiferromagnetic (AFM) phase transition (343 K) above room temperature[2]. At the same time, the widely popular $PbZr_{0.53}Ti_{0.47}O_3$ (PZT) thin films have also been extensively studied for potential and practical applications in dynamic and non-volatile ferroelectric (FE) random access memories due to their large remanent polarization ($P_r$), small coercive field and high Curie temperature[3,4]. We combine the basic properties of these two materials -- the AFM phase transition 343 K of PFW and the FE phase transition (~ 620 K) of PZT -- to make a series of innovative complex single-phase films and bulk ceramics $Pb(Zr_{0.53}Ti_{0.47})_{1-x}(Fe_{0.66}W_{0.33})_xO_3$ ($PZTFW_x$) (x= 0.20 (a), 0.30 (b), 0.40 (c)) at ambient temperature using chemical solution technique (CSD), PLD, and sintering.

At present there are only two well-known room-temperature magnetoelectrics -- $BiFeO_3$ (BFO) which is multiferroic with ($P_r$ ~55 μC/$cm^2$ @ 15 kHz, $M_s$ ~ 1 emu/$cm^3$)[5,6], and $Cr_2O_3$, which is not multiferroic The magnetoelectricity in $Cr_2O_3$ was theoretically predicted by Dzyaloshinskii[7] and experimentally observed by Astrov[8]. Very recently Levstik et al. [9] have discovered magnetoelectric relaxors: that is, crystals with nanoregions of short-range order, with simultaneous frequency dispersion of both the electrical and magnetic susceptibilities. In search of other such prospective multiferroics we found a single-phase $Pb(Fe_{0.66}W_{0.33})O_3$(PFW)-$PbTiO_3$ (80:20) solid solution that showed electric dipole and magnetic spin order near room temperature with high loss only above its phase transition temperature[10]. We show in the present investigation that similar structures with PZT instead of $PbTiO_3$ (in particular, PZTFW with x=0.20) permit magnetic control of large polarization at room temperature[11,12,13].

There are few single-phase materials in nature that posses both ferroelectric and ferromagnetic properties independently[1]. Multiferroics may be either single-phase materials[5-11] or artificially designed nanostructuctures[14,15] with different ferroic orders such as (anti)ferroelectricity, (anti)ferromagnetism, and ferroelasticity coexist simultaneously. Recently multiferroics have generated attention due to the presence of two or more order parameters and the coupling among them. Physicists and engineers are fascinated by the physics and fabrication of multiferroic materials due to their potential applications in highly sensitive actuators and sensors, multi-state memories, and electronics[5,6,9,10,,14,15].

In the present scenario a new single-phase multiferroic is needed that can exhibit multiferroicity at room temperature. No reports of the processing of $PZTFW_x$ as polycrystalline bulk ceramics, single-crystals, or thin films have previously been published. From our initial studies of $PZTFW_x$ polycrystalline thin films we found that by tailoring the compositions, exploiting strain effects, and pressure, we can make a promising multiferroic at ambient temperature. In this report we investigate the dielectric constant and loss, polarization, and magnetic properties of $PZTFW_x$ (x= 0.20 to 0.40) thin films at ambient temperature. The goal of this study is to formulate innovative single-phase multiferroic thin films near room temperature with tuning of compositions. Recently we have shown the magnetic control of polarization in PZTFWx (x=0.20) which also possesses very high resistivity, can be explained by the fourth order $E^2H^2$ terms[11,13] that are indirect via strain (electrostriction plus magnetostriction).

A Venn diagram (Fig.1) helps design new novel room temperature magnetoelectric (ME) multiferroics which consist of the most popular ferroelectric (PZT) and three weak relaxor-type multiferroics: PFW, PFN and PFT[16,17,18,2]. Since all four belong to the Pb(B`B``)$O_3$ family, their solid solutions are all feasible. Films of a particularly novel single-phase complex perovskite PZTFWx ($0.20 < x < 0.40$) are synthesized here by different techniques; all showed room temperature ME multiferroicity.

## 2. Experimental details:

### Sol-gel films:

Ferroelectric PZTFW (x= 0.20 (a), 0.30 (b), 0.40 (c)) thin films were deposited on Pt/Ti/$SiO_2$/Si(100) substrates using chemical solution deposition (CSD). The high purity (> 99.9 %) precursor materials (i.e., lead acetate trihydrate, iron 2-4 pentanedieonate, titanium isopropoxide, zirconium isopropoxide and tungsten isopropoxide) were brought from Alfa Aesar and used for making stock solutions. For good quality films a PZTFWx solution of 0.2-0.3 molar concentration was spin coated at 3000 rpm for 30 seconds and pyrolyzed at $400^0$C for 2 minutes. This process was repeated several times in order to get desired film thickness. Rapid thermal annealing (RTA) was performed at different temperatures from 600-$750^0$C for different time intervals ~ 60-300 seconds to get the desired phase and high density. DC sputtering was carried out for depositing the Pt top electrode of 3.1 x $10^{-4}$ $cm^2$ area using a shadow mask. The dielectric properties in the frequency range of 100 Hz to 1 MHz were studied using an impedance analyzer HP4294A (Agilent Technology Inc.) over a wide range of temperature attached to a temperature controlled probe station (MMR Technology). Magnetic properties were investigated using a vibrating sample magnetometer (VSM) (Lakeshore model 7400). The polarization versus electric field (P-E) hysteresis loop of the capacitor was measured using the RT 6000HVS ferroelectric tester (Radiant technology) operating in virtual ground mode.

**Pulsed laser deposition films (PLD):**

Stoichiometric ceramic targets of PZTFW (x= 0.20 (a), 0.30 (b), 0.40 (c)) with 10% excess of lead oxide (to compensate for loss of volatile Pb) were synthesized by a conventional solid-state route. Three different compositions of PZTFWx films with varying deposition temperature were fabricated by pulsed laser deposition (PLD) employing a KrF excimer laser (λ=240 nm). The conducting $La_{0.67}Sr_{0.33}CoO_3$ (LSCO) layer was grown at $600^o$ C under an oxygen pressure of 300mTorr, using a laser energy density of (1.8 J/$cm^2$) and repetition rate of 10 Hz, followed by normal cooling in oxygen atmosphere. LSCO was used as bottom electrode for all kind of electrical measurements. The PZTFWx layer was then deposited on the LSCO layer at 600$^{o}$C under an oxygen pressure of 200mTorr, using a laser energy density of (1.5 J/cm$^{2}$) and repetition rate of 10 Hz. After deposition, PZTFWx was annealed at 600$^{o}$C for 30 minutes in oxygen at a pressure of 300Torr. Finally, the films were cooled down to room temperature at a slow rate. The total thicknesses of the films were around 350 nm. The structural analysis was done with a Siemens D500 x-ray diffractometer (Cu Kα radiation) in a θ-2θ scan. Atomic force microscopy (Veeco-AFM-contact mode) was used to examine the morphology and the surface roughness. The film thickness was determined using an X-P-200 profilometer. The electrical and magnetic measurements were carried out with the same equipments as for CSD films.

## 3. Results and Discussion:

### 3.1 Structural, dielectric, polarization and magnetic behavior of CSD films

X-ray diffraction patterns (XRD) of the PZTFW (x= 0.20 (a), 0.30 (b), 0.40 (c)) thin films deposited on Pt/$TiO_2$/$SiO_2$/Si substrates at 400°C and post-annealed at 700°C for 60 seconds, continuously further annealed at 650°C for 180 seconds for better homogeneity and crystallinity, as shown in the Fig. 1. In order to get the desired phase, several temperatures were chosen for RTA; the compositions listed provide the best single-phase compound at each annealing temperature. The XRD analysis indicates that the films grown at 700 °C were single-phase polycrystalline in nature with less than 1% impurities (pyrochlore) phases. The synthesized complex single-phase perovskite possesses more than five elements in the perovskite octahedra, which makes it very unlikely to get a 100% pure phase, but the electrical properties of all the films having less than 5 % pyrocholre were excellent, giving very similar polarizations and electrical responses. A preliminary XRD investigation was carried out with the well known POWD program[19] which fitted well a tetragonal crystal structure having lattice parameters a = 4.0217 Ǻ and c = 4.0525 +/- 0.0060 for PZTFW (x=0.20), and pseudo-cubic i.e. a = 4.0117 Ǻ and c = 4.0115 +/- 0.0031; and a = 4.0067 Ǻ and c = 4.0065 +/- 0.0034 for PZTFW (x=0.30 and x=0.40) respectively. The intensity of the (100) peaks decreases and (110) peaks increases with increase in Fe and W percentage, which indicates that both atoms are substitutional for the Zr and Ti atoms for the PZT matrix. Enhancement in the composition of Fe and W leads to a shift in Bragg peaks

(2 Θ) to higher angles and confirms solid solubility. We cannot yet determine the detailed crystal structure of these films, because they are polycrystalline and quite thin, but XRD data suggests that 0.30<x<0.20 may be the morphotropic phase boundary; above this compositions it was pseudo-cubic/rhombohedral. The surface morphology of the 350 nm thin films investigated by AFM in contact mode over a 5 x 5 μm area and 20 nm heights (not shown here)[11] exhibited well-defined grains with an average size 50-150 nm; among these most of the grain diameters were in the range 40–50 nm with some agglomeration of larger grains. Surface topography images of 5 x 5 μm regions revealed average surface roughness of 5-7 nm. The observed bigger grain size and higher surface roughness may be due to growth at high temperature and utilization of conventional CSD synthesis processes[20].

The dielectric constant and dielectric loss measured as function of temperature, frequency and compositions are shown in Fig. 2(a) and Fig. 2(b) respectively. Fig. 2(a) shows a broad dielectric dispersion from 100 Hz to 1 MHz over a wide range of temperatures for higher frequencies. The results obtained from Fig. 2 (a) and Fig. 2(b) are as follows: (i) high dielectric constants for all compositions, which increases with increase in Zr and Ti compositions, i.e. $PZTFW_x$ (x=0.20) (a) to $PZTFW_x$ (x=0.40) (c); (ii) Dielectric constant maximum temperature ($410^0$C to $220^0$C) shifted to lower temperatures with decrease in Zr and Ti compositions; (iii) very broad temperature dependence of the dielectric anomaly, which becomes step-like with increase in Zr and Ti compositions; (iv) temperature dependent tunability for $PZTFW_x$ (x=0.20) (a) ( i.e.~ 25-35 % (depending on frequency)) for 0.25<x< 0.40; (v) very low dielectric loss (3% to 8%) over a wide range of temperature; (vi) more frequency-dependent dielectric loss with decrease in Zr and Ti concentrations.

A modified Curie-Wiess law[21,22] provides a very elegant technique for predicting the values of the critical exponents and related parameters -- i.e., γ = degree of relaxation; Δ = full width at half maxima of the dielectric spectra. In a normal ferroelectric γ = 1, whereas in a typical relaxor, γ = 2. The diffuseness of ferroelectric phase transition can be evaluated with the help of the modified Curie-Weiss law. The modified Curie-Weiss law is as follows:

$$\frac{1}{\varepsilon(\omega,T)} = \frac{1}{\varepsilon_{\max}(\omega,T)}\left\{1+\left[T-T_{\max}\right]^{\gamma}/2\Delta^{2}\right\} \qquad (1)$$

where γ (1< γ <2) corresponds to degree of relaxation; Δ, broadening parameter; ε, dielectric constant; $\varepsilon_{max}$, dielectric constant maximum; $T_{max}$; temperature of dielectric maximum. Figure.2(c) shows the graph between $[(\varepsilon/\varepsilon_{max}) -1]$ and $(T-T_{max})$ at 10 kHz, which was observed to be linear. The value of the degree of relaxation obtained from the linear fit was computed to be γ ~1.78 (+/- 0.05), 2.00 (+/- 0.10), 2.00 (+/- 0.20) for 20 %, 30%, 40% PFW, respectively, which well matched that for a broad diffuse ferroelectric phase transition[21] above. If the linear fitting (modified Curie-Weiss law) of the experimental data points gives $\gamma > 2.00$, some extra parameters are needed for physically meaningful fitting. We included a fluctuation parameter related to domain wall motion that allows a better fit for these highly diffuse ferroelectric phase transitions with $\gamma \sim 2.00$ similar to that in relaxor ferroelectrics. Although all these compounds showed high frequency dispersion over wide range of temperature even far above the dielectric maximum temperature, like that in lead-free relaxor ferroelectrics, no temperature shift in dielectric maxima with increase in frequency was noticed (very broad) unlike conventional relaxor ferroelectrics, so it is probably premature to label these compounds as normal relaxors. Having short-range ordering of polarization and a broad dielectric response with temperature is not sufficient to be a relaxor; that can alternatively arise from spatial inhomogeneity, as first suggested by Smolenskii.

The high dielectric constant, low dielectric loss, and high diffusivity could be explained on the basis of a statistical compositional fluctuation (i.e., disorder in the arrangement of different ions on crystallographically equivalent sites) in Fe:W:Zr:Ti concentration, and/or the displacement of B-site ions (Ti) sitting at the center of the unit cell, which are respectively the main features of relaxor and normal ferroelectrics[10].

The electric-field-induced polarization switching (P-E) behavior was studied by Sawyer-Tower measurements. Figure 3 shows the ferroelectric hysteresis loops for $PZTFW_x$ ($x = 0.20$ (a), $x = 0.30$ (b), and $x = 0.40$ (c)) at 300 K, 200 K and 150 K respectively. The films exhibit well saturated hysteresis loops with remanent polarization ($P_r$) of about 22, 20, and 11 $\mu C/cm^2$ respectively for (a), (b) and (c) for 400 kV/cm external electric field. We observed little change in the coercive field with increase in applied electrical field. Enhancement in remanent polarization may be attributed to the relaxed local strain (due to larger grains) and/or greater compositional. In order to check the real-time polarization behavior of PZTFWx (x = 0.20) films and thus their utility in memory devices, we performed fatigue tests. Fatigue behavior of PZTFW thin films were carried out with 100 kHz bipolar square waves in the presence of 350 kV/cm external electric field. The first fatigue data points were obtained after $10^5$ cycles and continued until $10^9$ cycles, as shown in Figure 4. There was a modest polarization loss (<12%). Twelve percent decay in fatigue of PZTFWx (x= 0.20) thin films on a platinized silicon substrate after $10^9$ cycles much better than the early reports for PZT,[23] indicating its suitability for memory device applications. The high polarization and fatigue behavior of $PZTFW_x$ thin films may be explained on the basis of the presence of a few very small polar nano-regions (PNRs). These PNRs may originate from the solid solutions of ferroelectric PZT and relaxor PFW (with a Burns temperature ~620 K). We also performed Piezo Force Microscopy (PFM) on these films (not shown here) which revealed small nano-regions between the interfaces of grains. We believe these nano-regions may provide a link among the grains and yield better ferroelectric polarization. Although one may argue that this is an influence of topographical effects, more analysis is needed. It is worth mentioning that with decrease in Zr and Ti concentration, the long-range ferroelectric domains diminish and the short-range polar nano-regions become more prominent, which results in a decrease in remanent polarization and saturation of the ferroelectric hysteresis loop.

The magnetization versus applied field (M-H) response of the $PZTFW_x$ (x = 0.20 (a), x = 0.30 (b), and x = 0.40 (c)) thin films at room temperature are shown in Figure 4.

M-H data of the $PZTFW_x$ (a) showed canted antiferromagnetic (AFM) spin order and very weak ferromagnetism. The magnetization of $PZTFW_x$ films grew with increases in Fe and W percentage. Under the application of an external 1.0T magnetic field the M-H curve shows an unsaturated loop for $PZTFW_x$ (a) with very low saturation magnetization ($M_s$~ 0.48 emu/cm$^3$) which further improve for x = 0.30 with ($M_s$~ 2.3 emu/cm$^3$), however for x = 0.40, we observed a well saturated weak ferromagnetic hysteresis with $M_s$~ 4.5 emu/cm$^3$. It was proposed that a super-exchange in the disordered regions through $Fe^{+3}$-O-$Fe^{+3}$ is expected to yield antiferromagnetic ordering[9,24]. Usually angles in the $Fe^{+3}$-O-$Fe^{+3}$ bonds are close to $180^0$. The compressive strain in the films may cause a small distortion in the $Fe^{+3}$-O-$Fe^{+3}$ spins. We believe that the observed weak room-temperature ferromagnetism is due to spin canting. On the other hand, all these complex compounds contains four ions $Fe^{+3}$, $W^{+6}$, $Ti^{+4}$, and $Zr^{+4}$ at the octahedral site of perovskite, which may provide an extra degree of freedom from $Fe^{3+}$ ($d^5$) ions to $Fe^{2+}$($d^6$) to the super-exchange in the disordered regions, which in turn could generate a better room temperature ferromagnetism with increase in Fe concentrations.

The magnetic behavior of polarization in these films has been described in our earlier paper[11]. It is related to the recent characterization of multiferroic relaxors by Levstik et al.[9] and, peripherally, to the model of magnetic-relaxor/ferroelectric-relaxor coupling of Shvartsman, Kleemann et al.[25] Pirc et al.[13] have shown that these results can be derived analytically from the spherical random-bond random-field (SRBRF) model and that the key term is biquadratic $E^2H^2$.

### 3.2 Structural, dielectric, polarization and magnetic behavior of PLD films

Figure 9 shows the XRD patterns of PZTFWx films along (100) directions. The peak positions were shifted towards higher Bragg angles with increase in Fe and W compositions. The full width at half maximum (FWHM) increases slightly for higher PFW percentages, indicating the local strain contribution to the FWHM is reduced by the presence of a unique crystalline orientation. Cell parameters of these films were estimated due to the limited number of peaks in XRD spectra. The details of crystal structure and their cell parameters are given in Table 1. Lattice mismatch between the substrate and the films were analyzed on the basis of the in-plane epitaxial relationship PZTFWx [100] || LSCO [100] || MgO [100]; we calculate the lattice mismatch using the equation $\varepsilon = \frac{a_{sub} - a_{film}}{a_{sub}} x100$ Where $\varepsilon$ is the lattice mismatch, $a_{sub}$ is the lattice constant of the substrate and, $a_{film}$ is the lattice constant of the PZTFWx. We observed a tensile strain (4.5-4.9% along the a-axis; 2.3-2.9% along the c-axis) and compressive strain (-4.9 to -4.5% along the a-axis, -7.3 to -6.6% along the c-axis) between the bottom electrode and the dielectric PZTFWx layer. Two oppositely-directed strains across the substrate-film and film-bottom electrode interfaces suggest qualitatively that in-plane compressive gradient strain (inhomogeneous strain) exists in these films, which may cause higher ferroelectric phase transition temperatures in PLD grown thin films, compared with the CSD specimens.

Surface morphology of a device is the key for any real practical applications. The surface morphology of the films was investigated by AFM in contact mode over 1μm x 1μm x 50 nm z-scale area, which indicates well-defined grains with an average size 40-100 nm; most of the grain size was in the range of 40–50 nm with some agglomeration of larger grains as shown in Fig. 10. The average surface roughness of all the compositions lies between 3-6 nm. The average surface roughness of the films decreases with increase in PFW/PZT ratio. Fig. 10 (c) reveals a darker reddish area with low z-height (< 10 nm) prominent in 40:60% FeW/TiZr compositions. It also suggests the presence of ordered regions -- i.e. polar nano regions (PNRs) -- in a disordered matrix. These PNRs contribute to the enhancement in dielectric dispersion and decrease in polarization. Some of the fine grains agglomerate due to higher fabrication temperature, which in turn gives the bigger grains seen in the AFM picture. The observed larger grain size and higher surface roughness may be due to growth at higher temperature and utilization of the high energy PLD process. PZTFWx (x =0.20) films showed well distributed homogenously packed grains throughout the surface; for higher PFW compositions more inhomogenous surfaces were observed.

The temperature and frequency dependence of PZTFWx for x = 0.20, 0.30, and 0.40 are shown in Figs. 11-13, respectively. Figure 11 shows frequency-independent dielectric peaks ($\varepsilon_{max}$) at 550 K from 100 Hz to 10 kHz and a 40 K shift in $\varepsilon_{max}$ temperature for higher frequencies (up until 1 MHz). These observations reveal strain-induced relaxor ferroelectric properties. The dielectric loss shows the character typical of relaxor materials: loss increases with increase in applied frequency. The frequency dispersion in the dielectric constant does not follow that of classical relaxor materials, but instead is more similar to that of impurity-induced relaxor materials like $Ba_{1-x}Sr_xTiO_3$ and $SrBi_2Nb_2O_9$. $SrBi_2Nb_2O_9$ can be tuned as a relaxor ferroelectric by doping with Ba, Sr, or Ce ions[26]. A huge frequency dispersion in the dielectric spectra below and above the ferroelectric phase transition is noticed with extra energy storage capacity during the heating and cooling process. Similar phenomena were observed in the loss-tangent spectra ε"(ω,T). Heating and cooling effects on the dielectric constant and loss spectra at selected frequencies (10 kHz and 100 KHz) are shown in Fig. 11(c) and (d).

Figure 12 shows the permittivity variation of PZTFWx (x= 0.30) as a function of temperature at various frequencies. It also follow the same rule as that of the previous compositions except for the following details: (i) huge shift (~100 K) in dielectric maximum temperature towards lower temperatures; (ii) a higher degree of frequency dispersion in both dielectric constant and loss spectra; (iii) hysteretic shift in dielectric maximum temperature during cooling (60 K @ 10 KHz); (iv) extra energy storage capacity during the heating and cooling processes; (v) broad diffuse phase transition for higher frequencies. Figure 13 shows the temperature dependent permittivity of PZTFWx (x= 0.40) as a function of temperature at various frequencies. This composition showed the highest frequency dispersion in dielectric below and above $T_c$. These observations suggest that more compositional disorder occurs at the B-site with increase in PFW/PZT ratio. A synopsis of dielectric constant and loss is given in Table. 2.

The most striking features of increased PFW/PZT ratio are as follows: (i) the ferroelectric phase transition temperature shifts towards room temperate; (ii) ferroelectric relaxor behavior (e.g., frequency dispersion) increases; (iii) energy storage capacity increases; (iv)more temperature hysteresis for $T_c$ upon heating and cooling; (v) loss tangent increases with increase in PFW /PZT ratio in solid solutions of PZTFWx.

The magnetization versus applied field (M-H) characterization of PZTFWx (x= 0.20) films performed for in-plane and out-of-plane field directions is shown in Fig. 14.

The induced magnetization values were between 6-8 emu/cc at 4 kOe magnetic field. M(H) along the c-axis showed well saturated ferromagnetic hysteresis with 2-3 emu/cc remanent magnetization and 230 Oe coercive field. The in-plane M(H) curve displayed a small opening (hysteresis) $H_c = 250$ Oe coercive field. The origin of magnetism can be explained in the light of the presence of $d^n$ ions at B-site which may provide in-plane ferromagnetism. The magnetization of PZTFWx (x=0.20) is better and well defined compare to same composition prepared by CSD techniques; this may be due to inhomogeneous (gradient) compressive strain in the films which cause a small distortion in the $Fe^{+3}$-O-$Co^{+3}$ spin angle, producing room-temperature ferromagnetism via spin canting. Super-exchange in the disordered regions through $Fe^{+3}$-O-$Fe^{+3}$ is expected to yield antiferromagnetic ordering; there is a possibility of strong magneto-electric coupling, suggesting the converse effect of our earlier report – viz., electric fields E controlling the magnetic properties at both mesoscopic and microscopic scale[27].

Figure 14 shows the temperature dependent magnetization from 4-300 K for all the investigated materials at 100 Oe. An exponential enhancement in the magnetization was observed above 180 K, whereas non-linear magnetization was seen up until room temperature. The nonlinear behavior in magnetization and the well saturated hysteresis at room temperature suggest the presence of ambient ferro/antiferro-magnetism in all the compositions. These magnetizations increase with increase in the PFW/PZT ratios. Zero-field-cooled (ZFC) magnetization (not shown) of PZTFWx (x= 0.20) films also exhibited a cusp at 180 K, suggesting the presence of polar-nano magnetic regions in the films. Nevertheless, detailed magnetic characterizations of these materials are still unfinished.

### 3.3 Sintered bulk ceramics

M. Kosec, B. Malic et al. at the Jozef Stefan Institute in Ljubljana have successfully followed our CSD deposition scheme to fabricate sintered bulk ceramics. The characteristics of these bulk specimens closely resemble those of our thin films. Although their work will be published elsewhere, we include reference to their work to alert readers to the existence of such samples for other measurements.

## 4. Summary

In conclusion, a novel class of room-temperature single-phase multiferroic $PZTFW_x$ ($0.20<x<0.40$) thin films has been synthesized by CSD and PLD techniques. A new and large family of room-temperature magnetoelectric mutiferroic devices can be designed at by tailoring compositions of PZTFWx at the B-site. It is proposed that with suitable compositions, we can make both ferroelectric and ferromagnetic phase transition temperatures coincide near room temperature, which will be of great interest for devices and basic physics. Ferroelectricity diminished whereas relaxor behavior and anti/ferromagnetism improved with increase in FeW/ZrTi ratios. High dielectric constant, low dielectric loss, diffused ferroelectric phase transition, low leakage current and high temperature dependent dielectric tunability were observed for all compositions. The temperature of the dielectric maximum temperature shifted strongly to lower temperatures with increase in Fe and W percentages. The ferroelectric phase transition temperatures in PLD films are higher than that of CSD specimens with the same nominal compositions, most likely attributable to in-plane gradient compressive strain. The ferro/antiferromagnetic/magnetic relaxor properties of PLD films are better that those CSD films. The basic properties of these new multiferroics promise a bright future for room-temperature magnetoelectrics having both ferroic phase transitions at ambient temperature, despite the fact that their magnetoelectric coupling is nonlinear..

**Acknowledgements:**

This work was partially supported by DOD W911NF-05-1-0340, W911NF-06-1-0030, W911NF-06-1-018, DoE FG 02-08ER46526 grants to UPR, and EU STREP Multiceral funding at Cambridge. We thank Barbara Malic and Marija Kosec for unpublished data on their sintered ceramics and Robert Blinc and Rasa Pirc for helpful discussions.

**Parameters for pulsed laser deposited PLD films of $Pb(Fe_{2/3}W_{1/3})_x(Zr_{0.48}Ti_{0.52})_{1-x}O_3$**

**Table 1.**

| Characteristic | 20%PFW/80%PZT | 30%PFW/70%PZT | 40%PFW/60%PZT |
|---|---|---|---|
| structure | Tetragonal (nm)<br>a = 0.4022±0.0006<br>c = 0.4115±0.0006 | Tetragonal (nm)<br>a = 0.4012±0.0003<br>c = 0.4103±0.0012 | Tetragonal (nm)<br>a = 0.4007±0.003<br>c = 0.4089±0.0014 |
| Electrical polarization | $P_r$ = 58 μC/cm$^2$ at 295K | $P_r$ = 33 μC/cm$^2$ at 295K | $P_r$ = 30 μC/cm$^2$ at 295K |
| Magnetization | $M_r$ = 2.11 emu/cc at 295K | $M_r$ = 2.29 emu/cc at 295K | $M_r$ = 4.53 emu/cc at 295K |
| Dielectric constant ε(295K) | 500 | 1292 | 1002 |
| Dielectric loss δ(295K~10 kHz ) | 0.058 | 0.079 | 0.091 |

**Table 2. Lattice mismatches strain in highly oriented PZTFWx thin films**

| Substrate/ sublayer | Lattice parameter (Å) | Lattice mismatch (%) with LSCO | Lattice mismatch (%) with PZTFWx (x=.20) | Lattice mismatch (%) with PZTFWx (x=.30) | Lattice mismatch (%) with PZTFWx (x=.40) |
|---|---|---|---|---|---|
| MgO rocksalt | 4.213 | 8.97 | 4.53<br>2.32 | 4.77<br>2.61 | 4.88<br>2.94 |
| LSCO cubic | 3.835 | ------------ | -4.85<br>-7.30 | -4.61<br>-6.98 | -4.48<br>-6.62 |

## Figure Captions

**Fig. 1** A Venn diagram is proposed to design new novel room temperature magnetoelectric (ME) multiferroics

**Fig. 2** Room temperature XRD of the PZTFW (x= 0.20 (a), 0.30 (b), 0.40 (c)) thin films with rapid thermal annealed at 700ºC for 60 sec than 650ºC for 180 sec. All these films have less than 1% pyrochore phase.

**Fig. 3** Temperature dependent **(a)** dielectric constant of sol-gel derived PZTFWx thin films at 10 kHz and 50 kHz over wide range of temperature

**Fig. 4** Temperature dependent tangent loss of sol-gel derived PZTFWx thin films at 10 kHz and 50 kHz over wide range of temperature

**Fig. 5** Dielectric diffusitivity [($\varepsilon/\varepsilon_{max}$-1) and (T-$T_{max}$)] at 10 kHz sol-gel derived PZTFWx thin films.

**Fig. 6** Polarization vs. electric field hysteresis loops for sol-gel derived $PZTFW_x$ (x = 0.20 (a), x = 0.30 (b), and x = 0.40 (c)) at 300 K, 200 K and 150 K respectively.

**Fig. 7** Fatigue behavior of the films with 100 kHz bipolar square waves in the presence of 350 kV/cm external electric field over $10^9$ cycles. P-E hysteresis is given over wide range of applied field (inset).

**Fig. 8** Room temperature magnetization versus magnetic field (M-H) response of the $PZTFW_x$ (x = 0.20 (a), x = 0.30 (b), and x = 0.40 (c)) thin films.

**Fig. 9** Room temperature XRD of the PZTFWx (x= 0.20 (a), 0.30 (b), 0.40 (c)) thin films grown by PLD technique.

**Fig. 10** Surface morphology of PLD grown PZTFWx (x = 0.20 (a), x = 0.30 (b), x = 0.40 (c),) thin films in contact mode over 1µmX1µm x 50 nm z-scale area.

**Fig. 11** Room temperature polarization vs. electric field hysteresis loops for PLD grown $PZTFW_x$ (x = 0.20, x = 0.25, x = 0.30, x = 0.40 with very high polarization, well saturation and low coercive field.

**Fig. 12** Temperature dependent PLD grown $PZTFW_x$ (x = 0.20), **(a)** dielectric constant vs. temp from 1kHz to 1MHz **(b)** dielectric constant vs. temp during heating and cooling at 10 kHz and 100 kHz **(c)** tangent loss vs. temp during heating and cooling at 10 kHz and 100 kHz, **(d)** tangent loss vs. temp from 1kHz to 1MHz; (clock wise).

**Fig. 13** Temperature dependent PLD grown $PZTFW_x$ (x = 0.30), **(a)** dielectric constant vs. temp from 1kHz to 1MHz **(b)** dielectric constant vs. temp during heating and cooling at 10 kHz **(c)** tangent loss vs. temp during heating and cooling at 10 kHz, **(d)** tangent loss vs. temp from 1kHz to 1MHz; (clock wise).

**Fig. 14** Temperature dependent PLD grown $PZTFW_x$ (x = 0.40), **(a)** dielectric constant vs. temp from 1kHz to 1MHz **(b)** dielectric constant vs. temp during heating and cooling at 10 kHz and 100 kHz **(c)** tangent loss vs. temp during heating and cooling at 10 kHz and 100 kHz, **(d)** tangent loss vs. temp from 1kHz to 1MHz; (clock wise).

**Fig. 15** Room temperature magnetization versus magnetic field (M-H) response of the PLD grown $PZTFW_x$ (x = 0.20) thin films (a) in plane magnetization, (b) out of plane i.e. magnetization along c-axis.

**Fig. 16** Temperature dependent magnetization for all the compositions from 4 K to 300 K.

[25] V. V. Shvartsman, S. Bedanta, P. Borisov, W. Kleemann, Phys. Rev. Letters, **101**, 165704, (2008).

[26] P. Lehnen, W. Kleemann, T. Woike, and R. Pankrath, Phys. Rev. B **64**, 224109, (2001)

[27] R. Blinc, P. Cevc, A. Zorko, J. Holc, M. Kosec, Z. Trontelj, J. Pirnat, N. Dalal, V. Ramachandran, and J. Krzystek J. Appl. Phys. **101**, 033901, (2007)

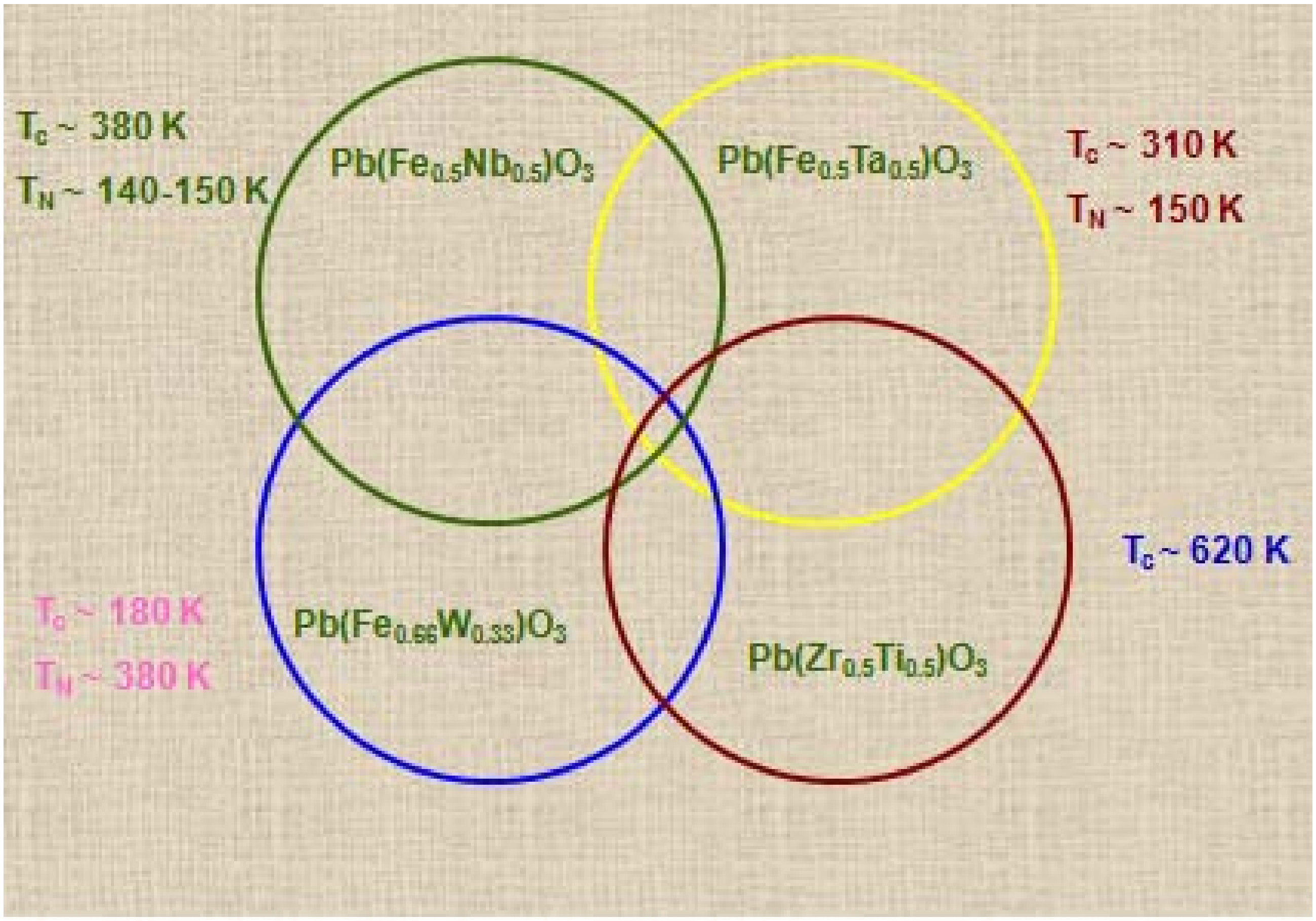

T_c ~ 380 K
T_N ~ 140-150 K
Pb(Fe0.5Nb0.5)O3
Pb(Fe0.5Ta0.5)O3
T_c ~ 310 K
T_N ~ 150 K
T_c ~ 620 K
T_c ~ 180 K
T_N ~ 380 K
Pb(Fe0.66W0.33)O3
Pb(Zr0.5Ti0.5)O3

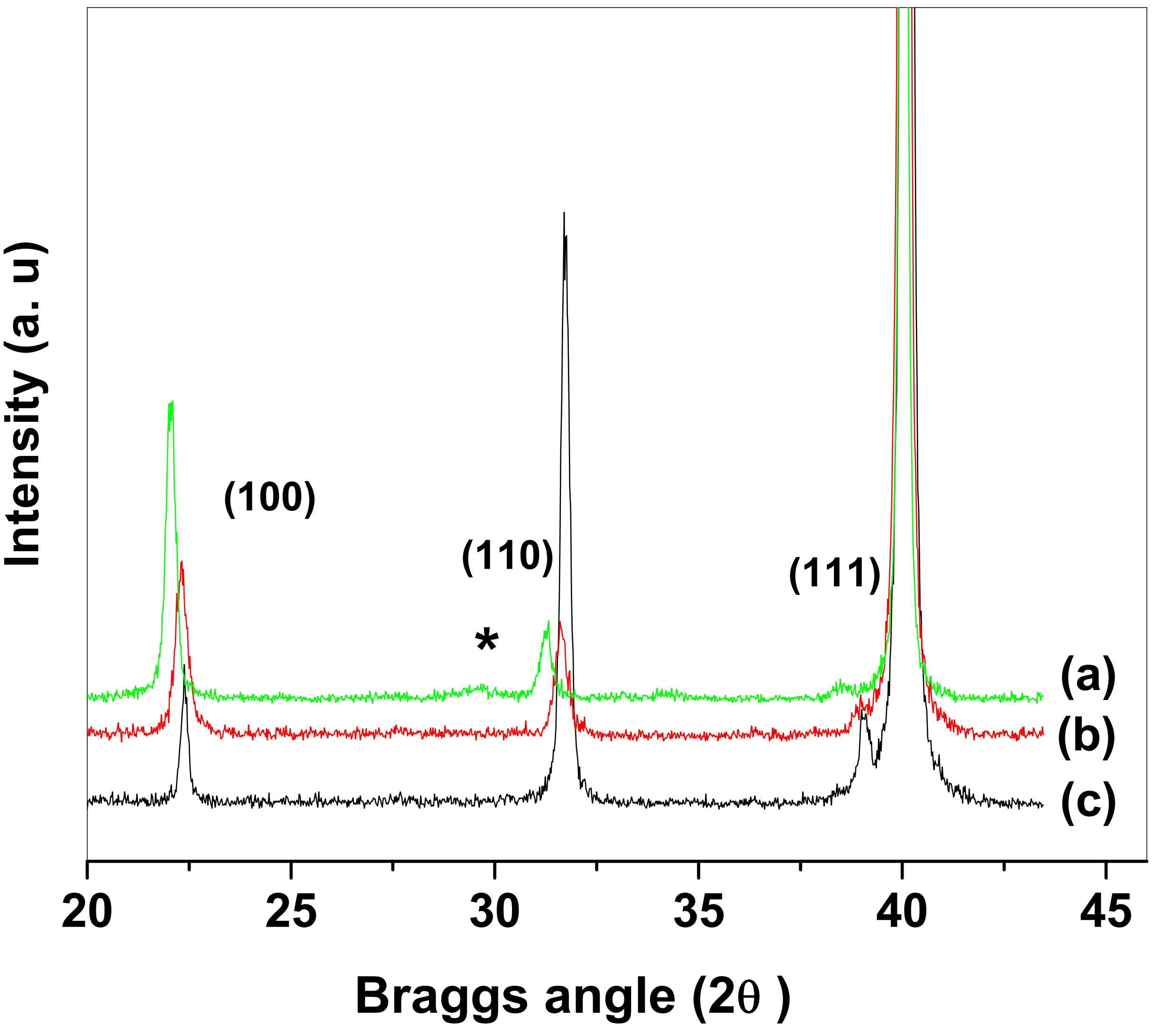

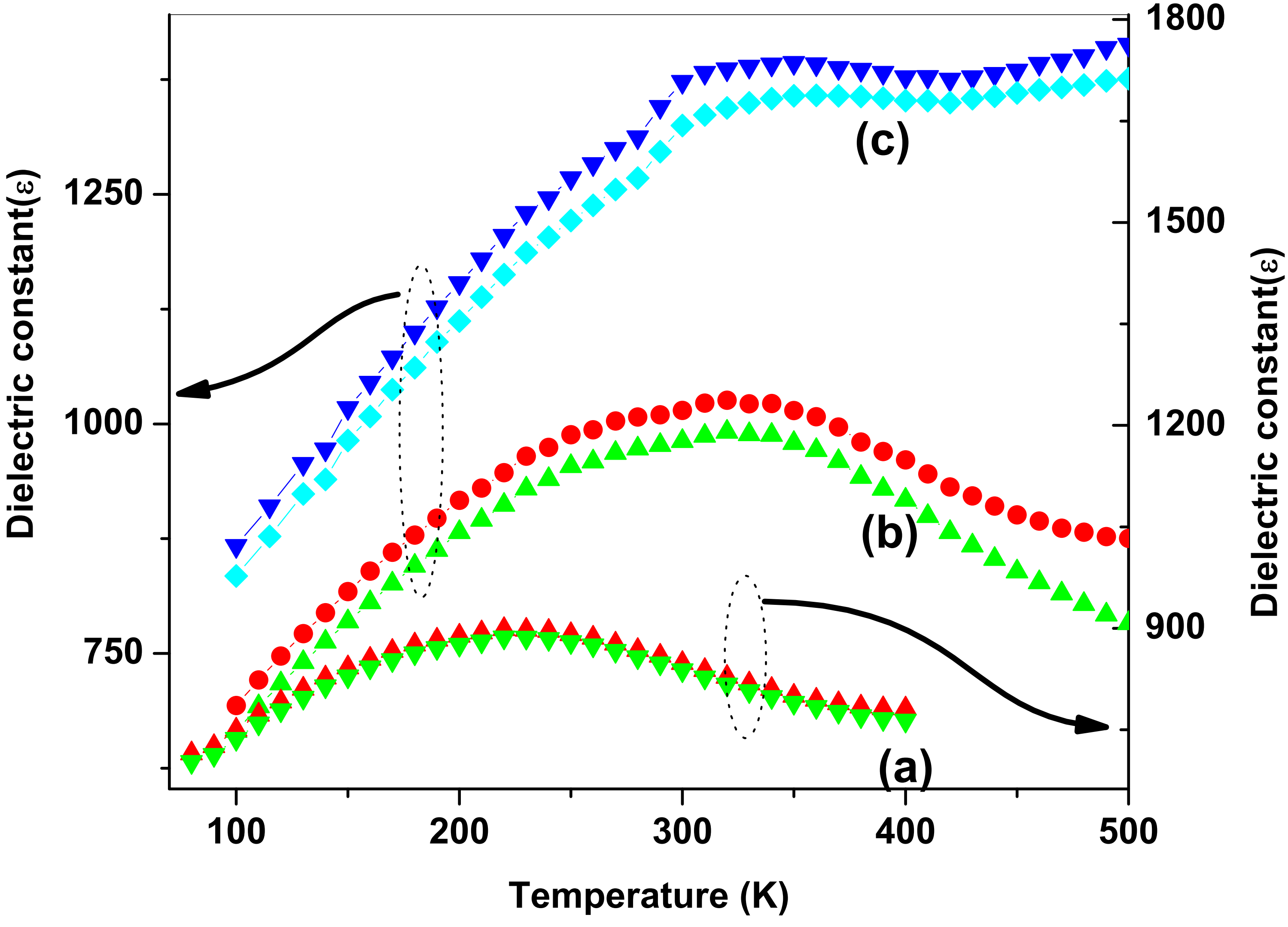

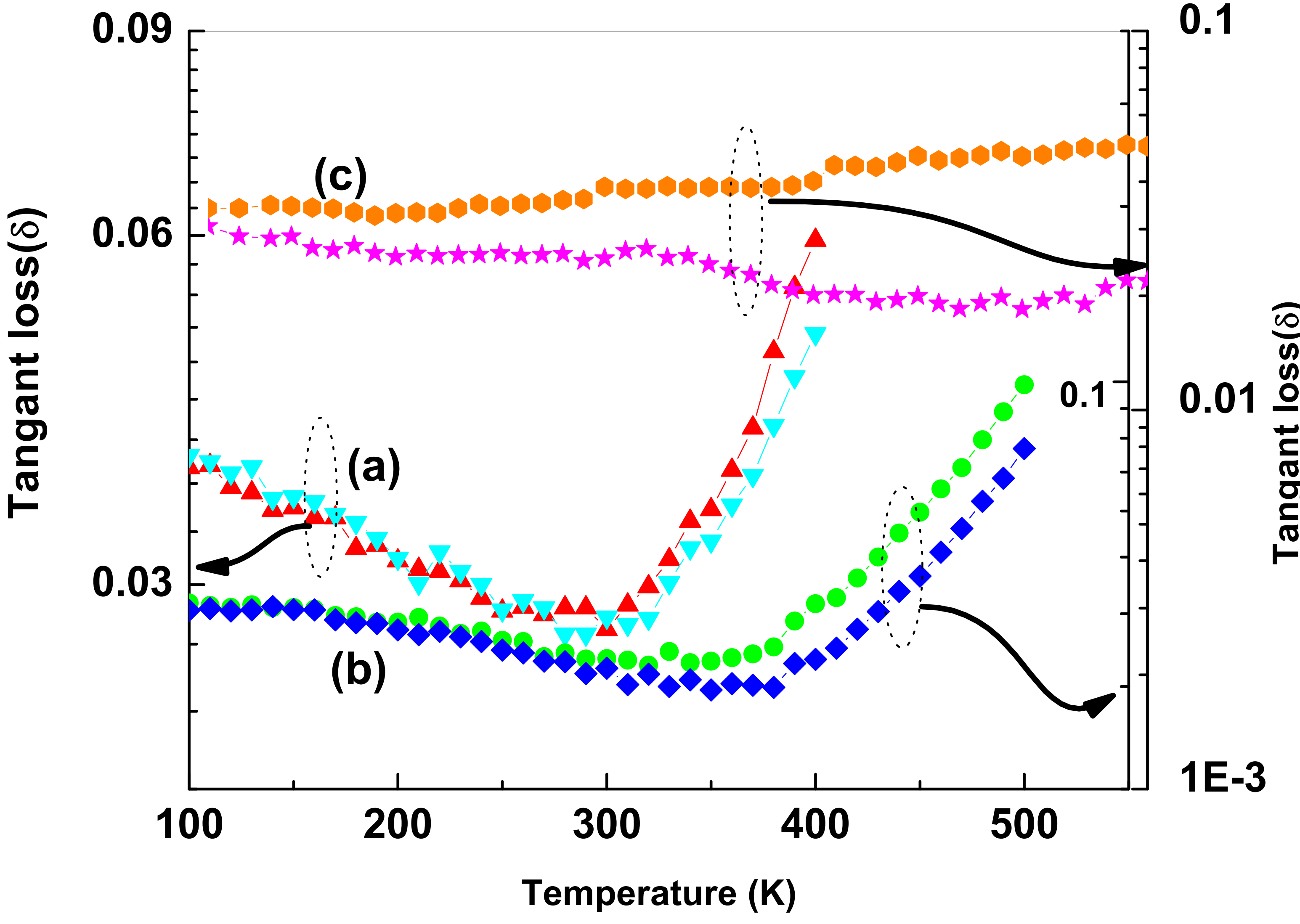

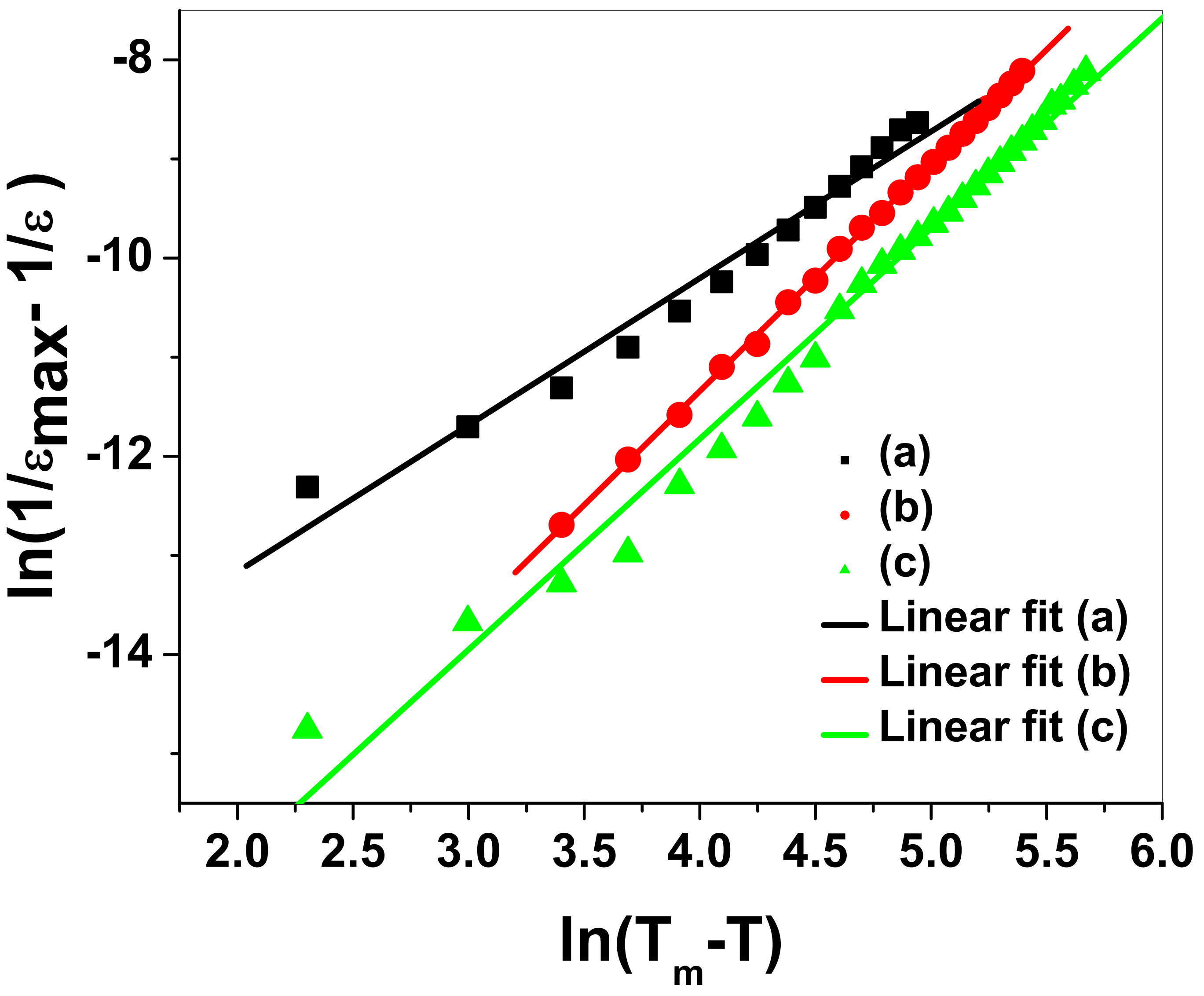

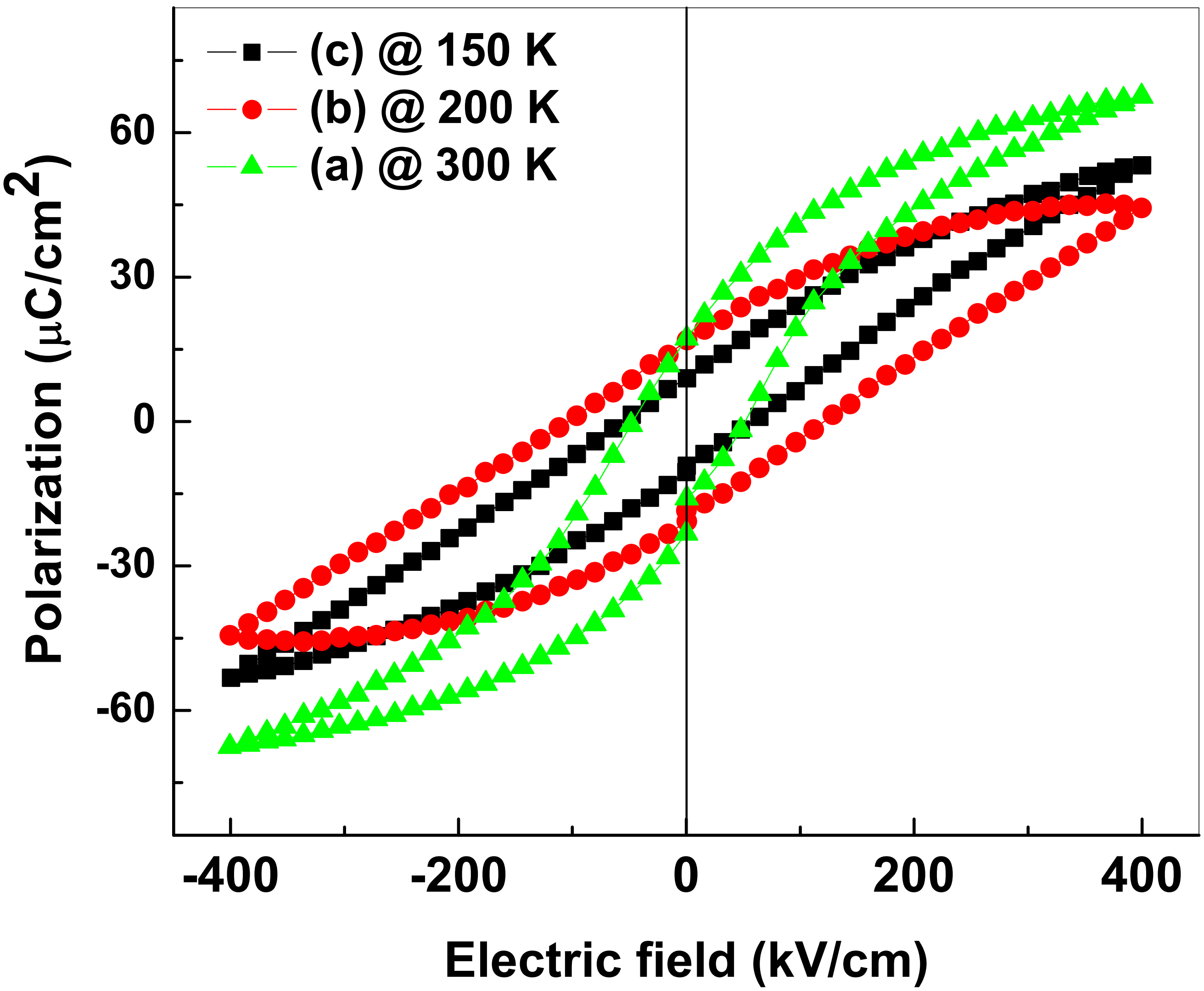

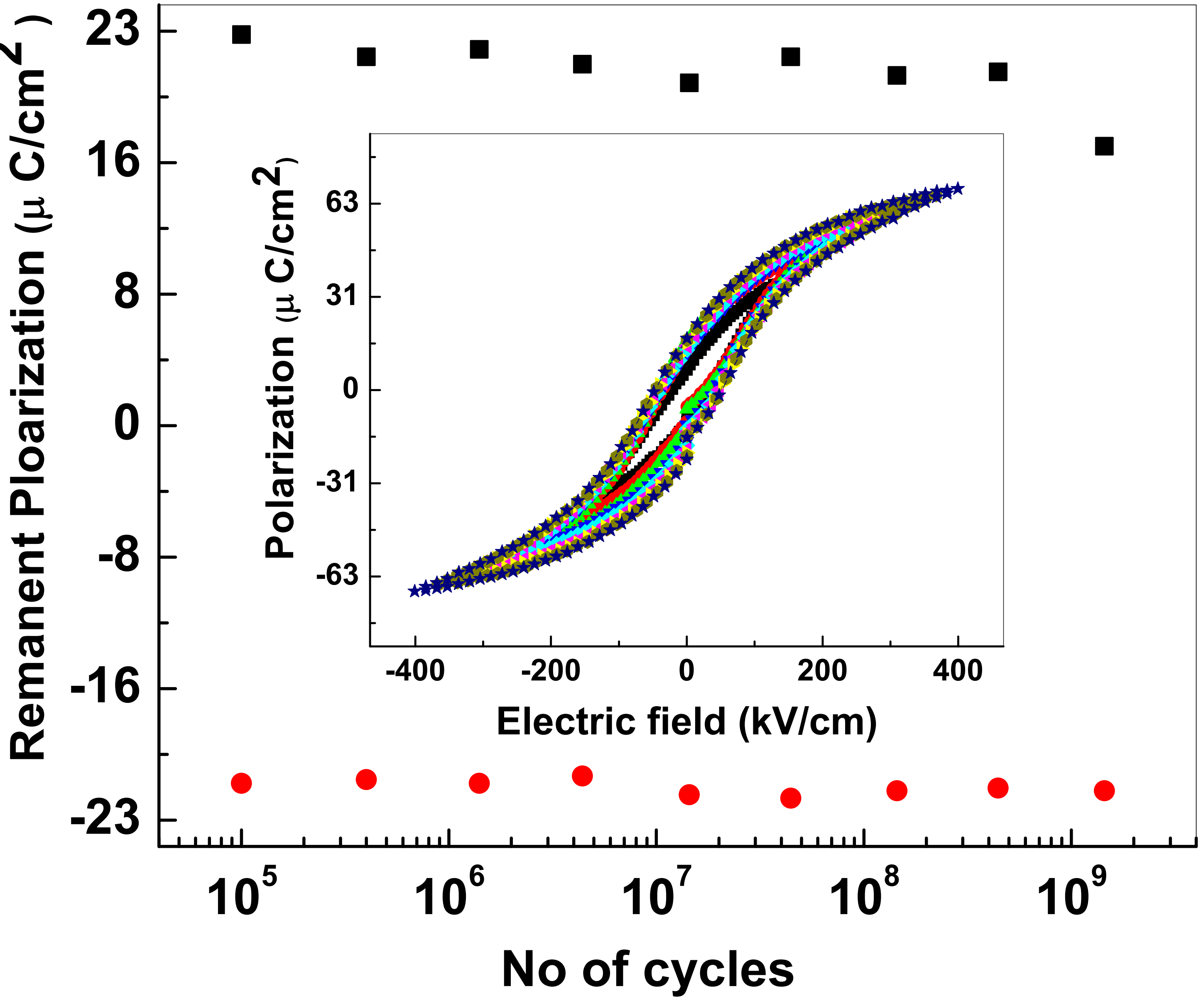

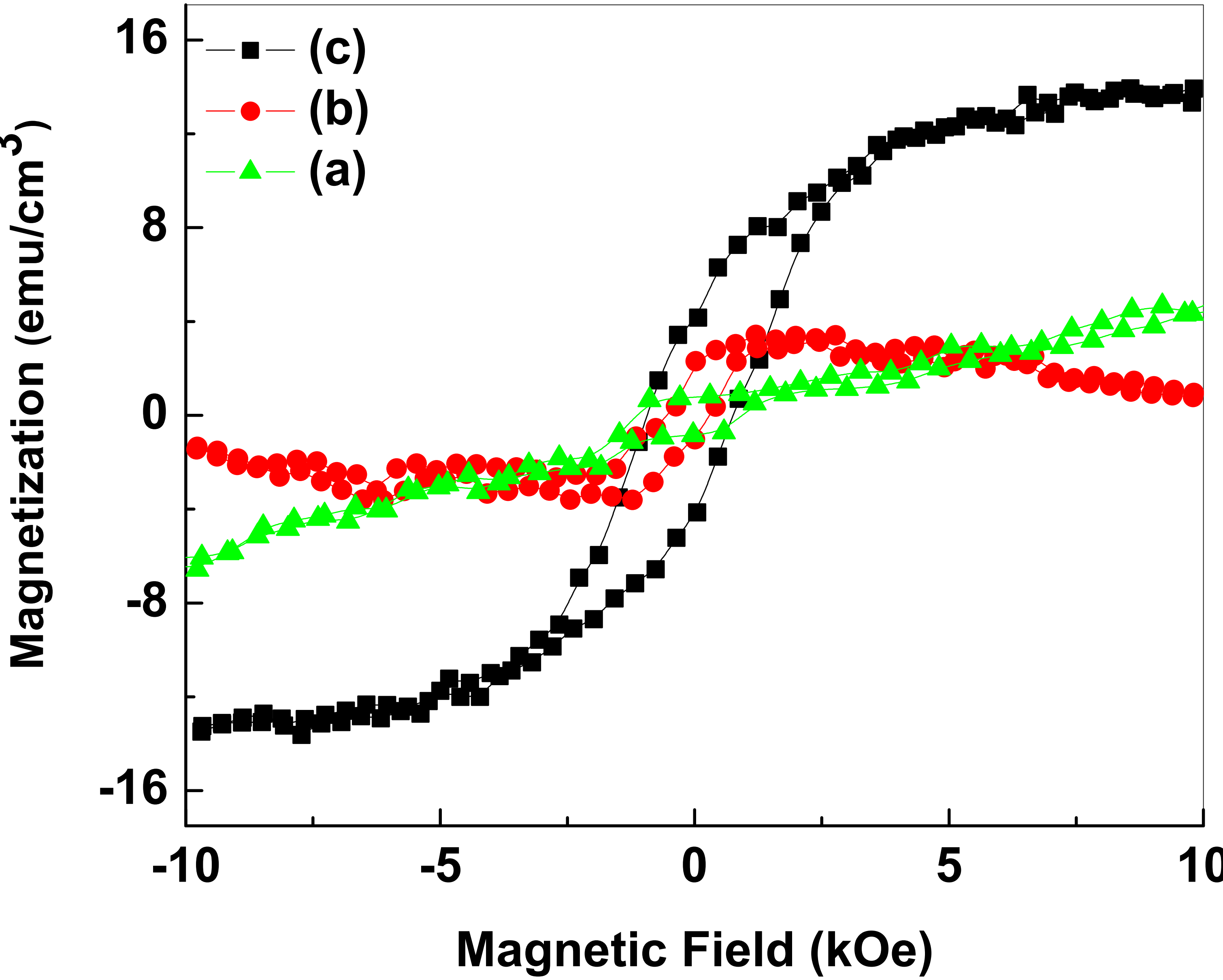

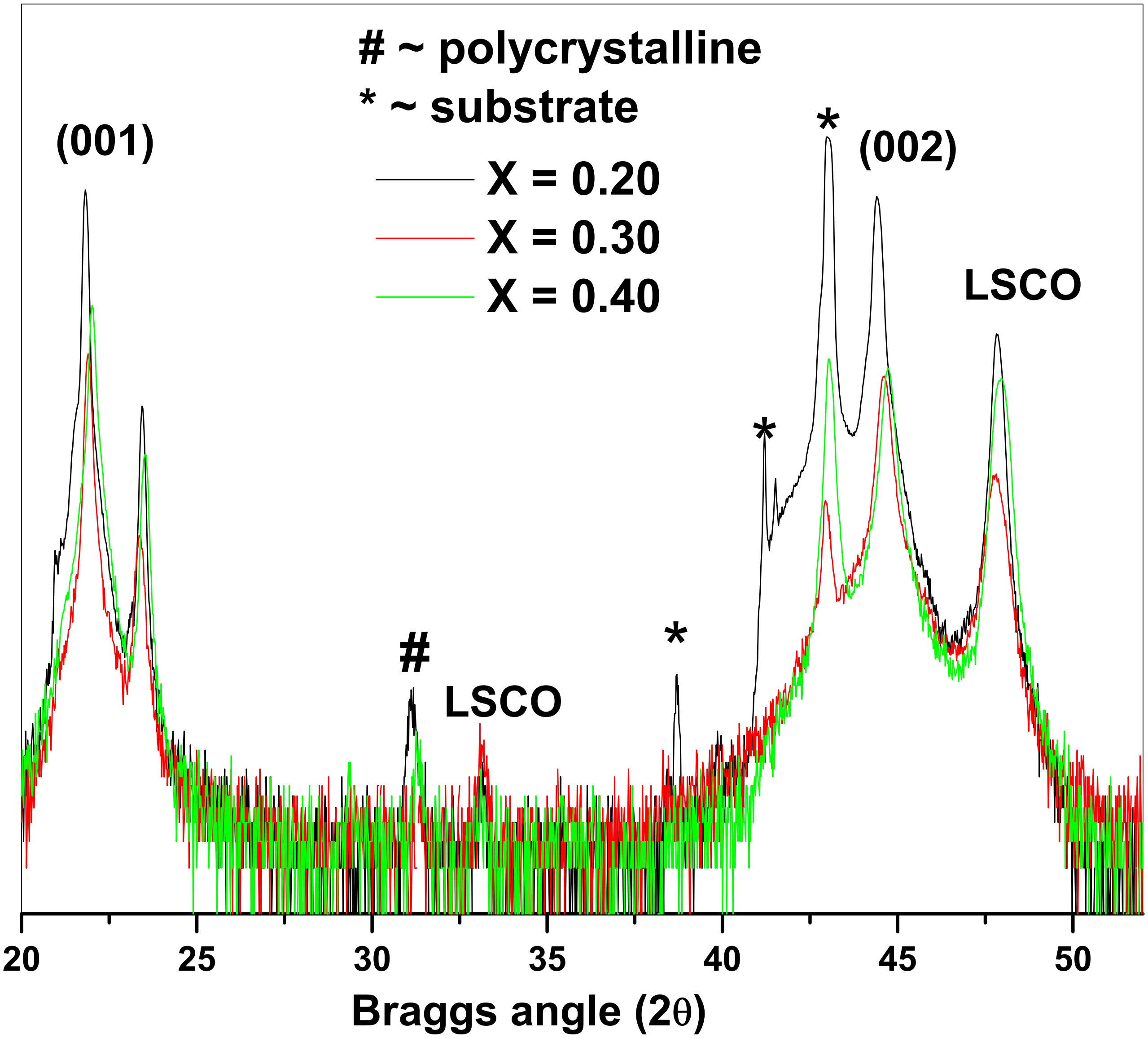

**(a)**

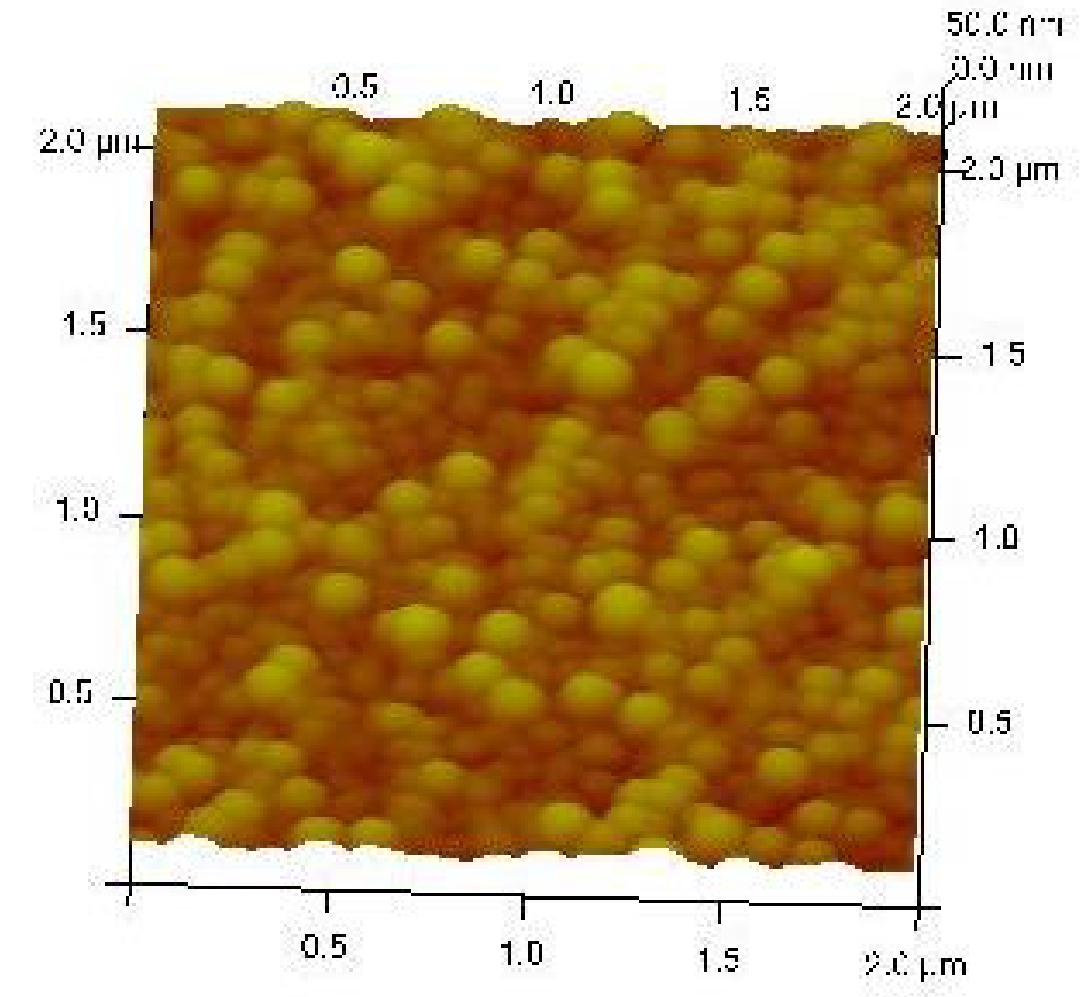

**(b)**

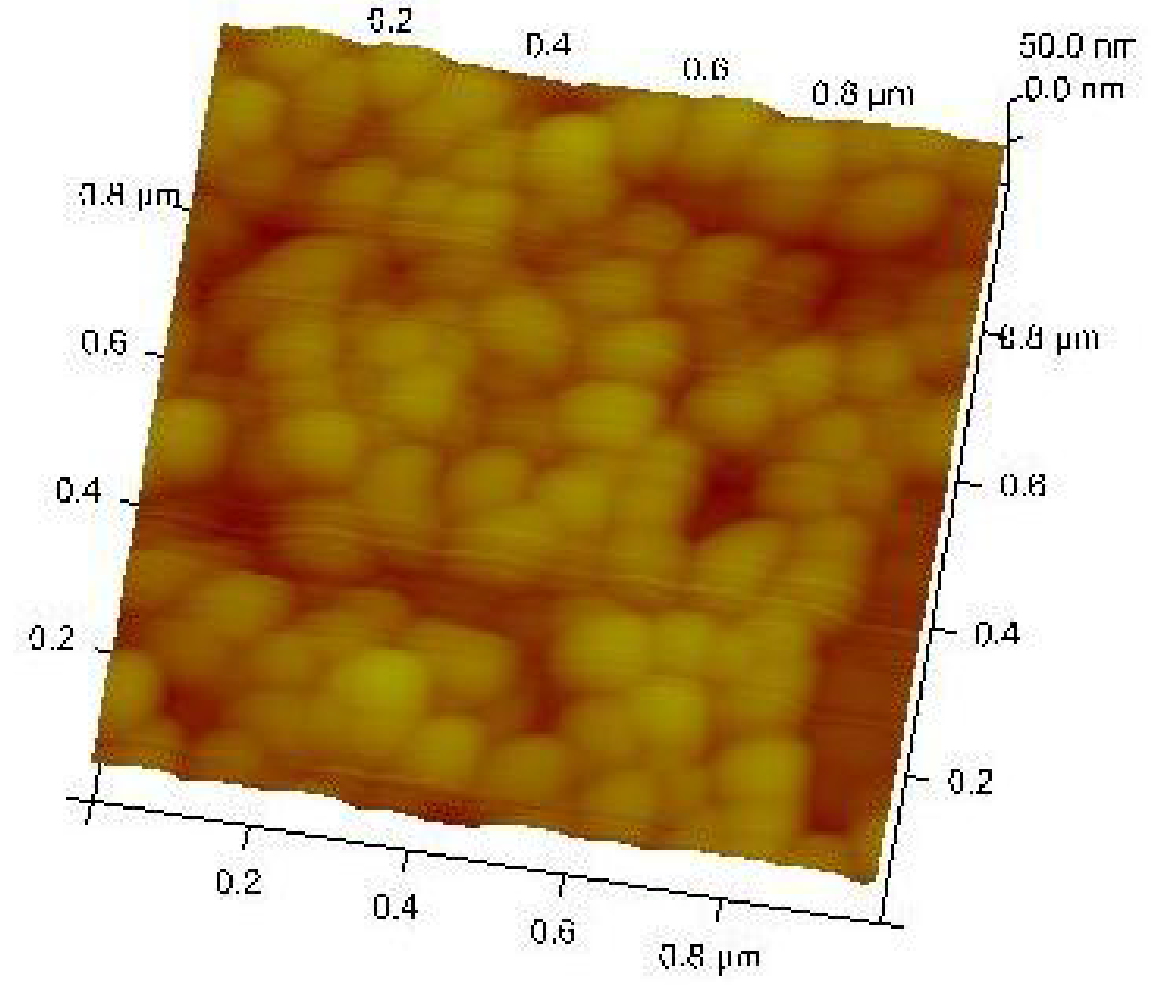

**(c)**

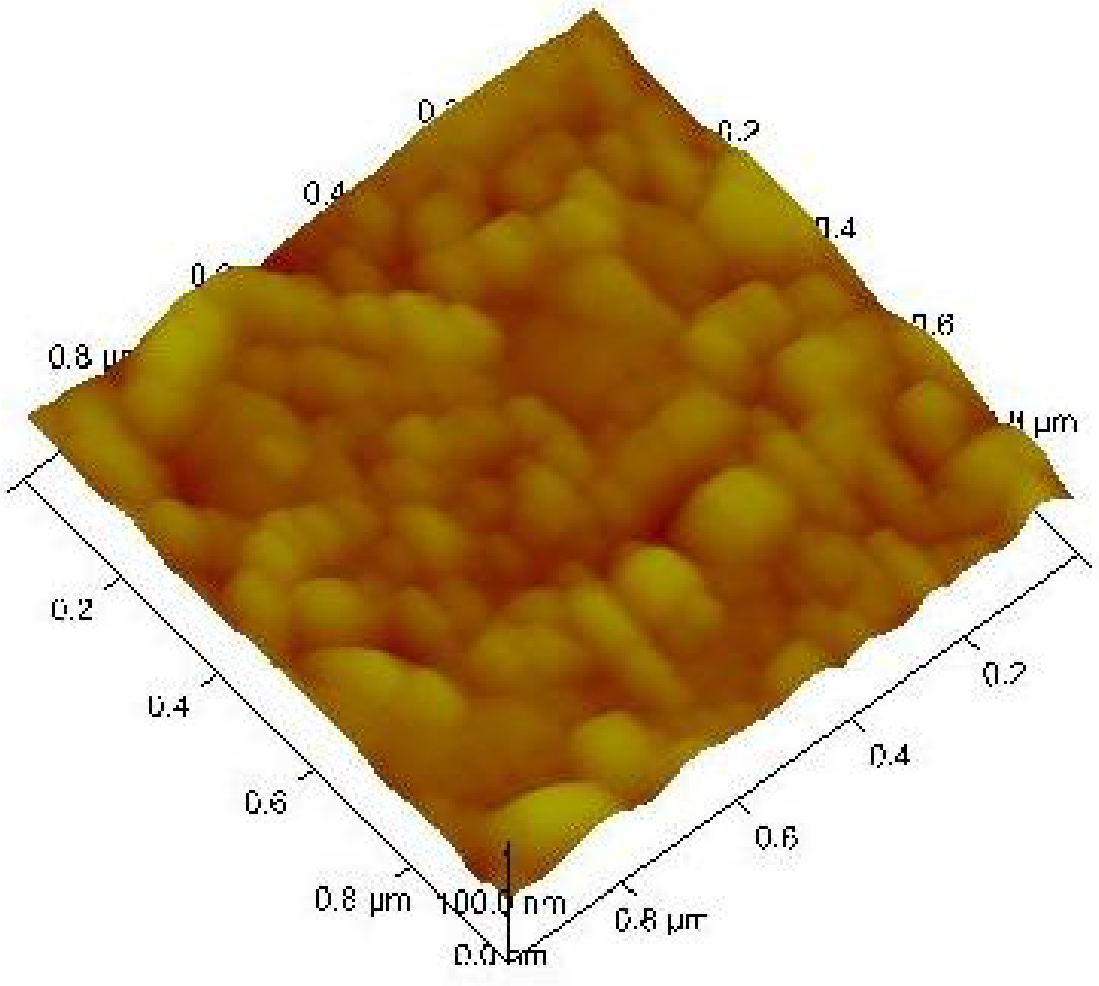

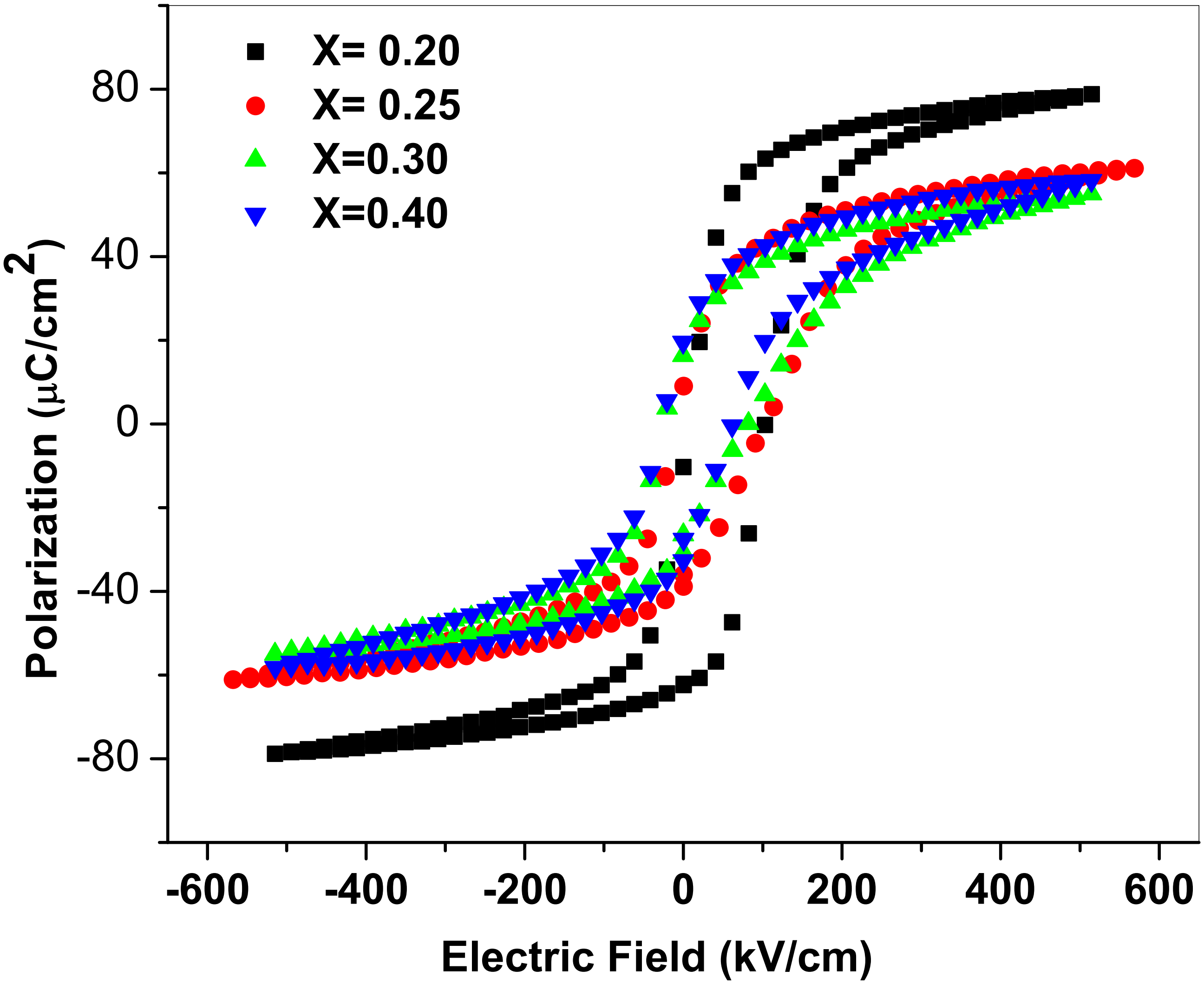

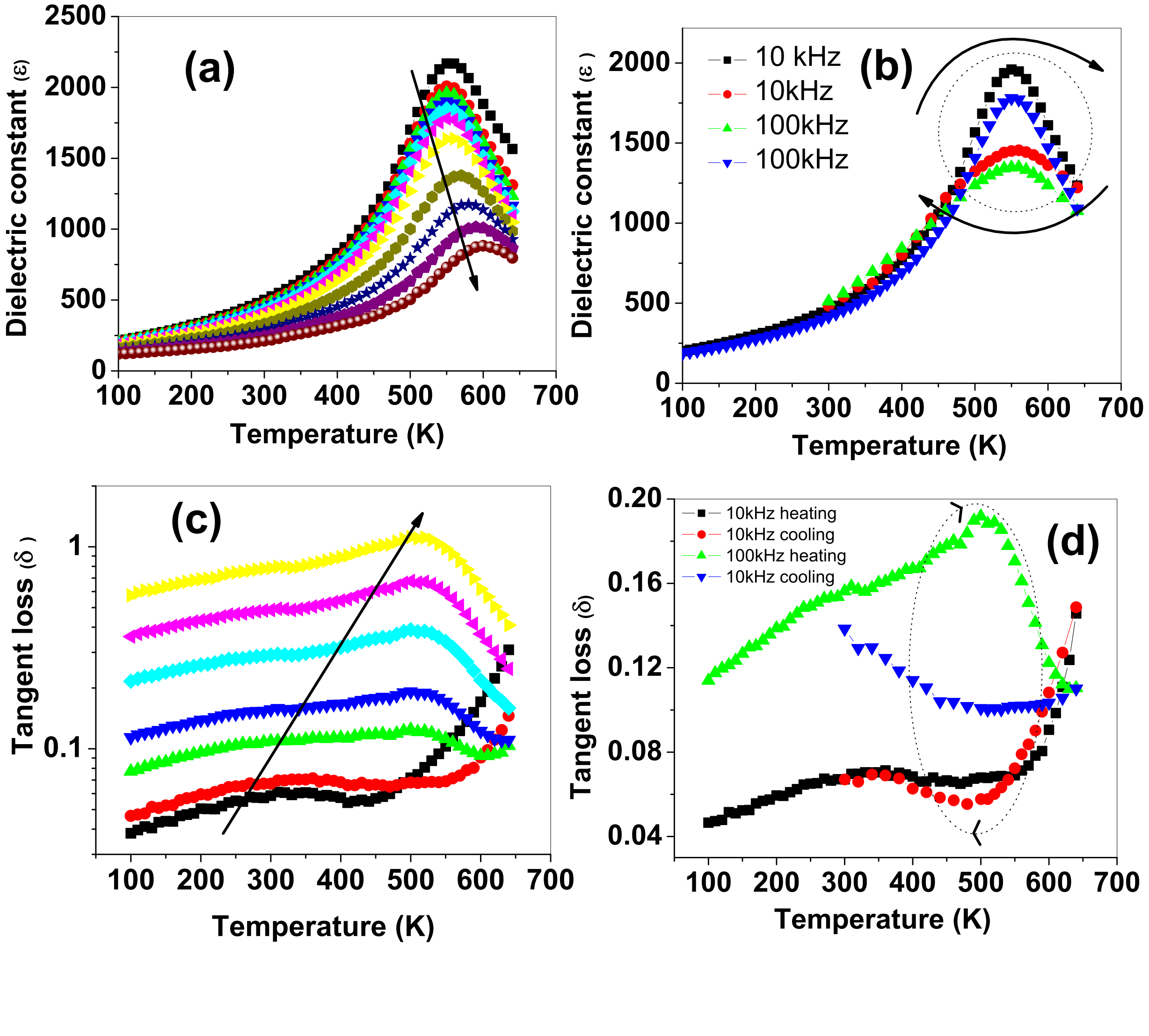

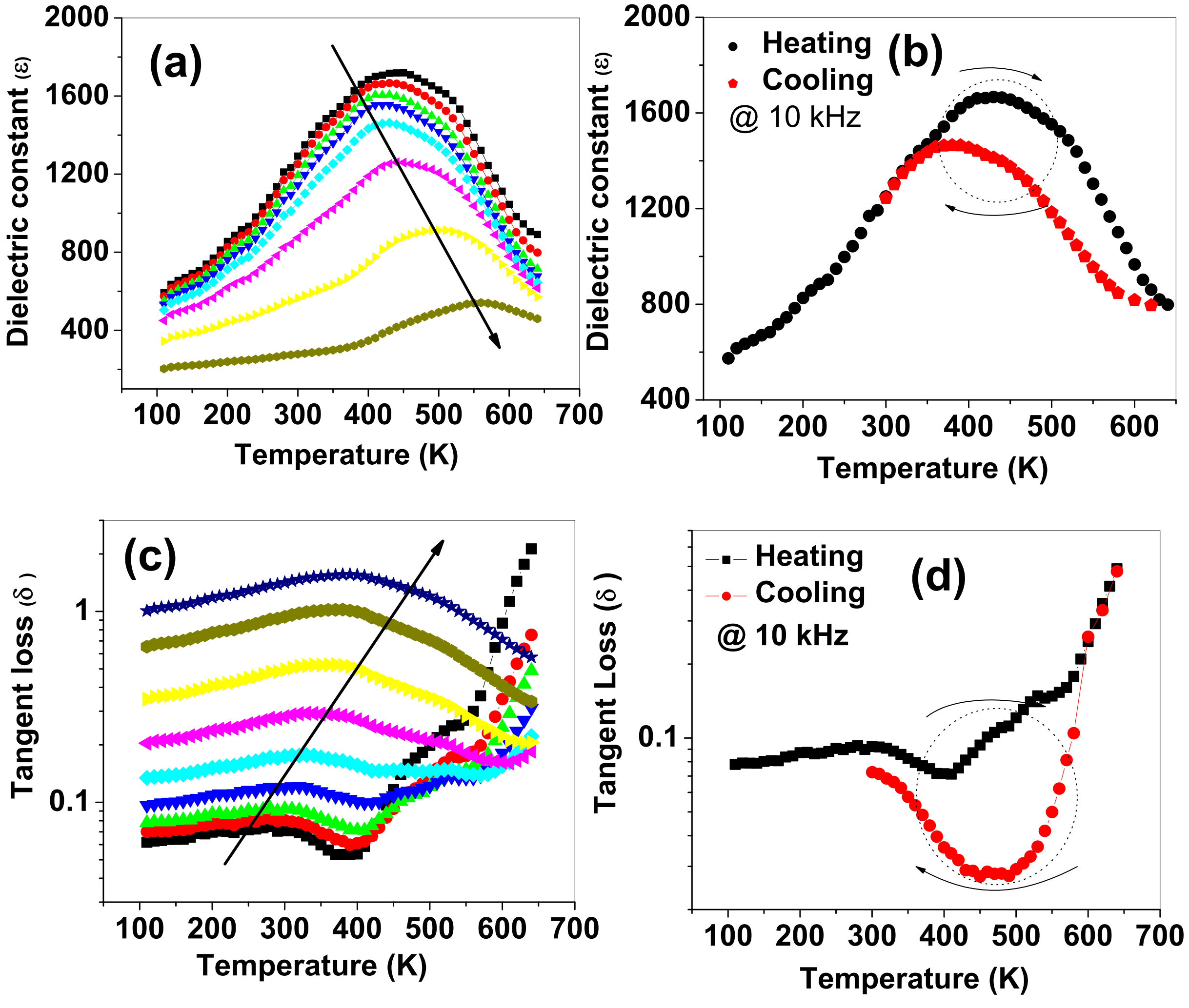

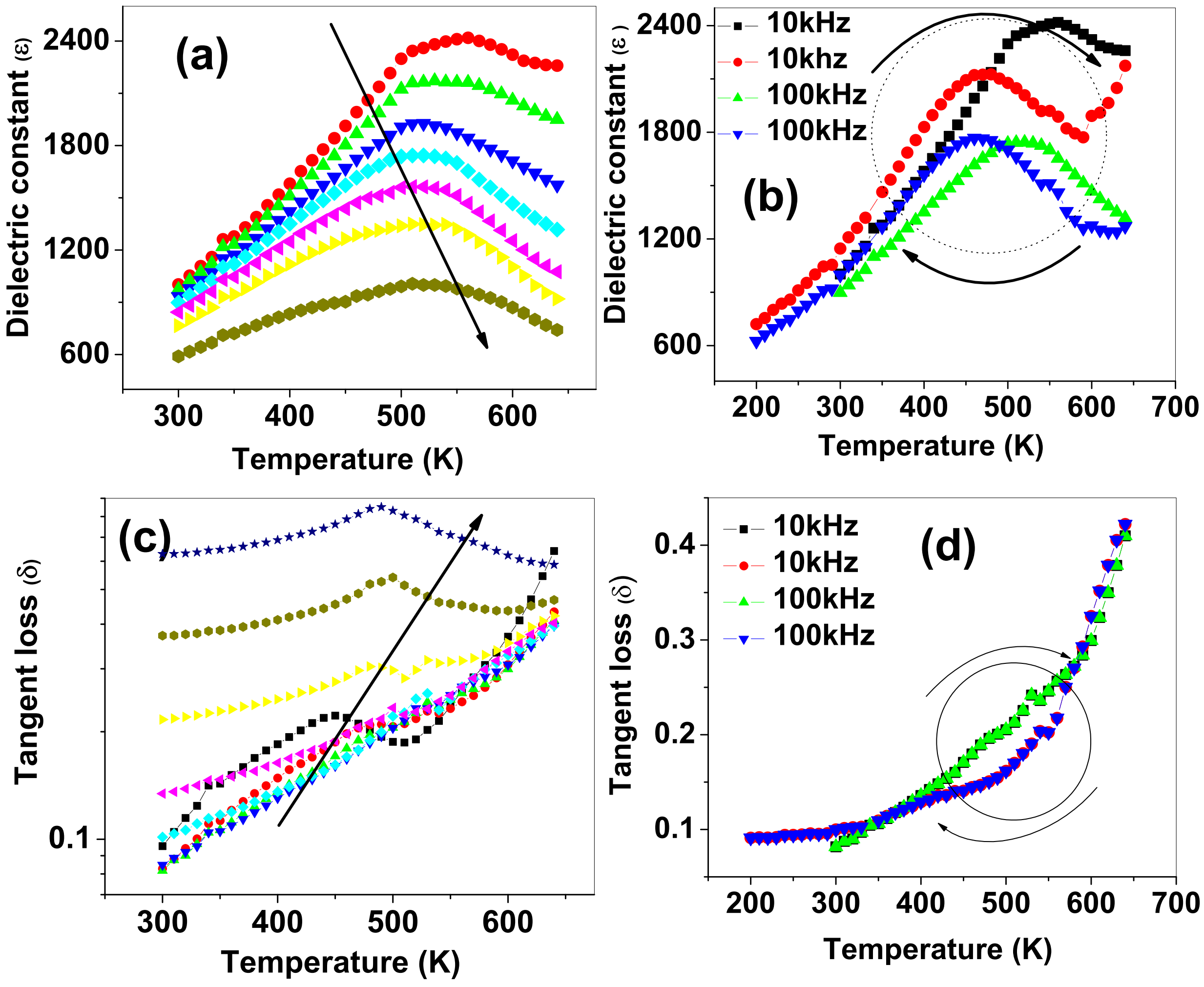

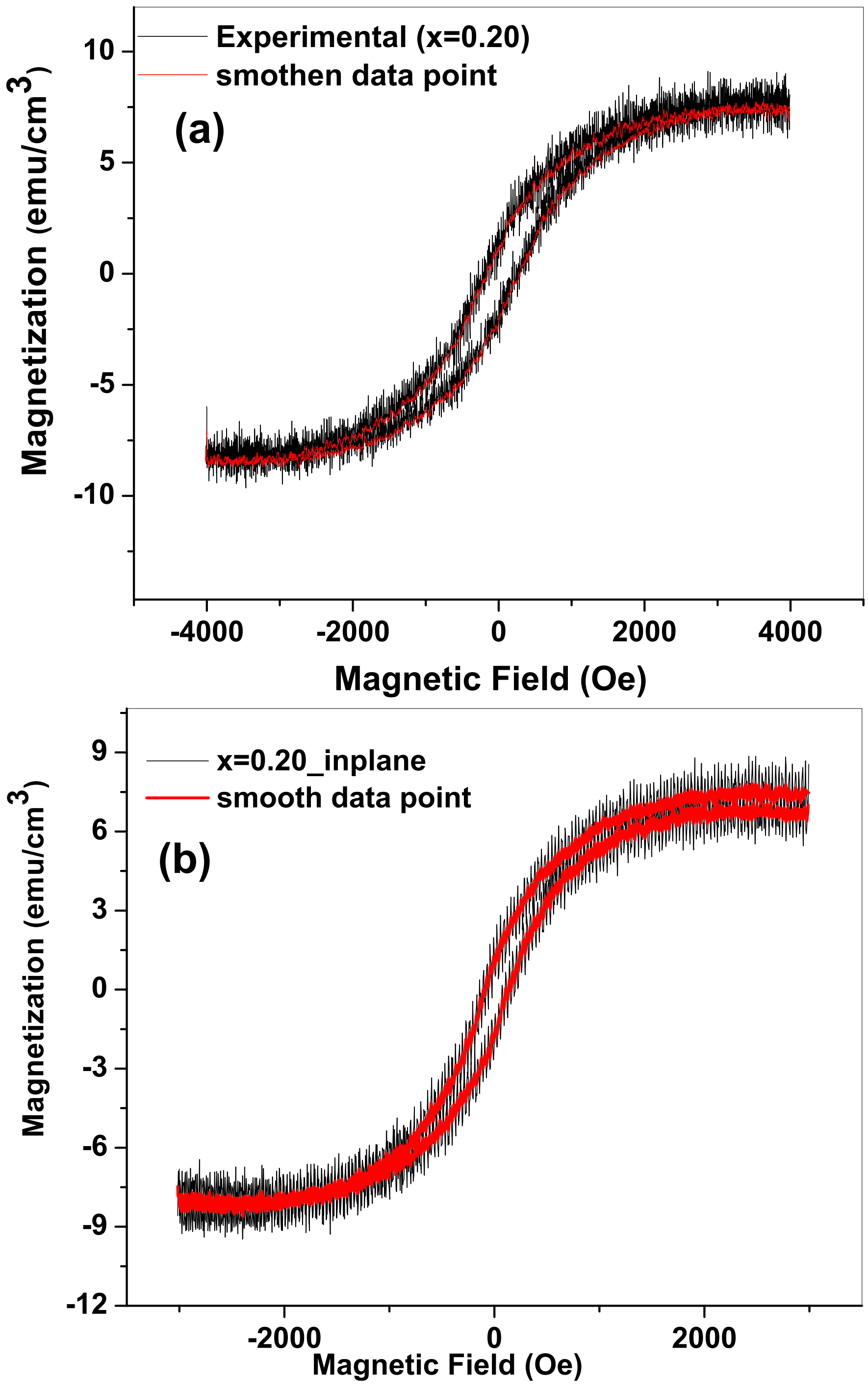

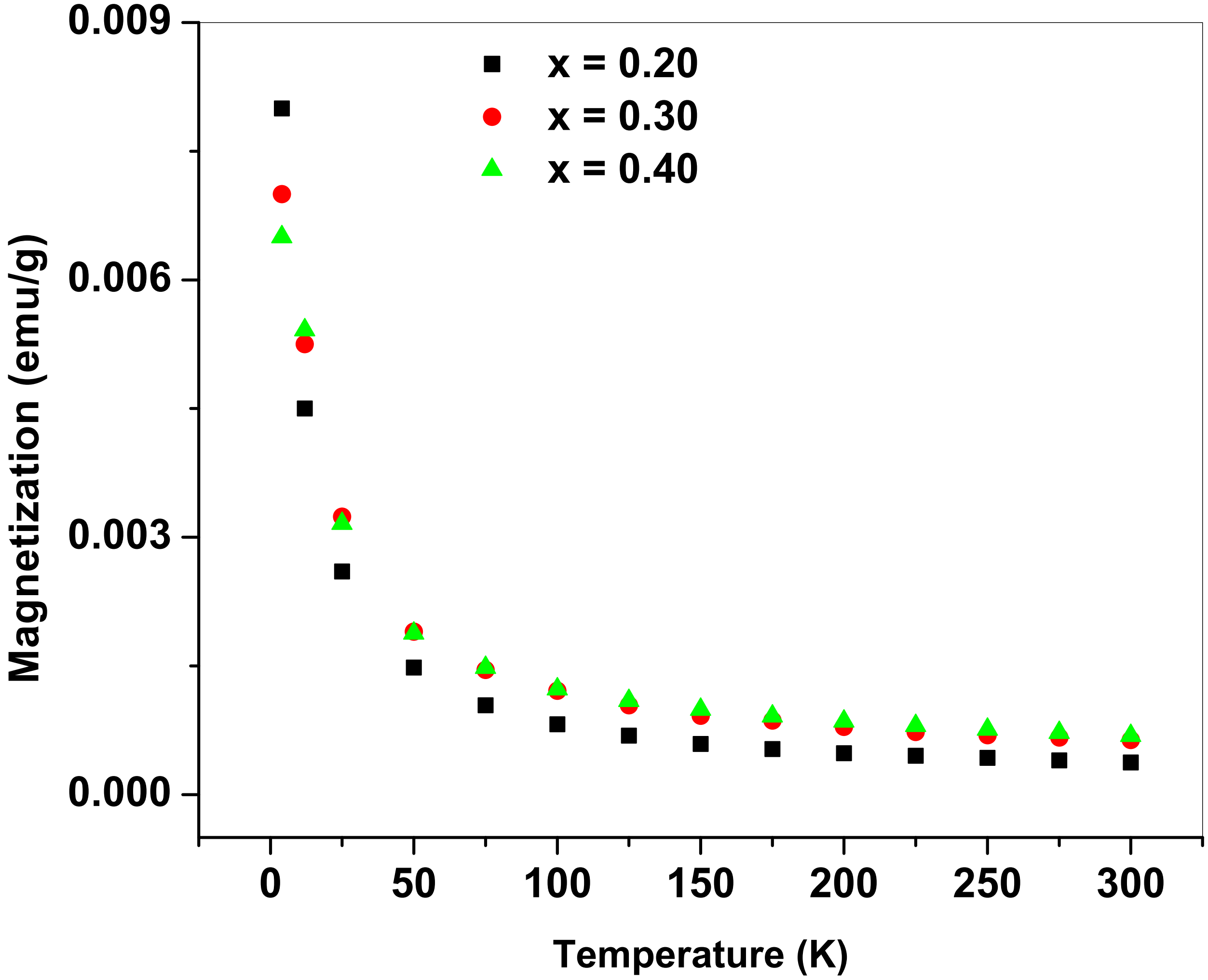